\documentclass[pre,aps,twocolumn,showpacs,preprintnumbers,superscriptaddress]{revtex4}

\usepackage{graphicx}
\usepackage{dcolumn}
\usepackage{bm}
\usepackage{amssymb}
\usepackage{pifont}
\usepackage{amsmath}
\usepackage{times}
\usepackage[nooneline]{subfigure}

\begin{document}
\title{Structure of characteristic Lyapunov vectors in spatiotemporal chaos}

\author{Diego Paz\'o}\email{pazo@ifca.unican.es}
\affiliation{Instituto de F\'{\i}sica de Cantabria (IFCA), CSIC--Universidad de
Cantabria, E-39005 Santander, Spain}

\author{Ivan G.\ Szendro}\email{szendro@ifca.unican.es}
\altaffiliation[Present address: ]{Max Planck Institute for the
Physics of Complex Systems, N\"{o}thnitzer Stra{\ss}e 38, 01187 Dresden,
Germany}
\affiliation{Instituto de F\'{\i}sica de Cantabria (IFCA), CSIC--Universidad de
Cantabria, E-39005 Santander, Spain}
\affiliation{Departamento de F{\'\i}sica Moderna, Universidad de
Cantabria, Avenida Los Castros, E-39005 Santander, Spain}

\author{Juan M. L{\'o}pez}\email{lopez@ifca.unican.es}
\affiliation{Instituto de F\'{\i}sica de Cantabria (IFCA), CSIC--Universidad de
Cantabria, E-39005 Santander, Spain}

\author{Miguel A. Rodr{\'\i}guez}\email{rodrigma@ifca.unican.es}
\affiliation{Instituto de F\'{\i}sica de Cantabria (IFCA), CSIC--Universidad de
Cantabria, E-39005 Santander, Spain}

\date{\today}

\begin{abstract}
We study Lyapunov vectors (LVs) corresponding to the largest Lyapunov exponents in systems
with spatiotemporal chaos. We focus on characteristic LVs and compare the results with
backward LVs obtained via successive Gram-Schmidt orthonormalizations. Systems of a very
different nature such as coupled-map lattices and the (continuous-time) Lorenz `96 model
exhibit the same features in quantitative and qualitative terms. Additionally we propose a
minimal stochastic model that reproduces the results for chaotic systems. Our work
supports the claims about universality of our earlier results~[I.~G. Szendro {\em et al.},
Phys.~Rev.~E {\bf 76}, 025202(R) (2007)] for a specific coupled-map lattice.
\end{abstract}


\maketitle

\section{Introduction}
\label{sec_intro}
Nonlinear spatially extended systems often exhibit spatiotemporal chaos (STC), {\it i.e.}
an apparent randomness in both space and time. Lyapunov exponents (LEs) measure the
exponential separation (or convergence) of nearby trajectories and provide an important
tool to characterize chaos in nonlinear dynamical systems~\cite{eckmann,bohr,ott}. Not
only exponential separation rates but also the associated directions in tangent space, the
so-called Lyapunov vectors (LVs), are required when trying to tackle many important
aspects of STC, such as, for instance, the role of hydrodynamic modes~\cite{yang08},
extensivity properties~\cite{egolf00} or predictability questions~\cite{bohr}, among
others. Random initial errors evolve in time and asymptotically align with the main LV
corresponding to the most unstable direction. In practice, this limit is reached
exponentially fast, so the memory of the initial perturbation is quickly lost.

In extended systems, the spatial distribution and correlations of LVs are crucial to deal
with questions such as predictability~\cite{primo05}. The relevance of spatial
correlations is particularly apparent in the context of weather forecasting (see for
instance~\cite{primo07}). 

Localization of LVs in several distributed systems has been noticed and discussed in some
extent in the
literature~\cite{pomeau84,kaneko86,giaco91,falcioni91,chate93,pik94,morriss03,morriss07}.
This phenomenon has been termed dynamical localization of errors~\cite{pik98}: The main LV
rapidly tends to concentrate around a narrow region of space. In homogenous systems, where
all degrees of freedom are equivalent, the localization locus is not anchored to any fixed
site, but keeps moving all over the system. However, in the case of inhomogeneous systems
LVs become strongly localized at certain fixed pinning centers and the phenomenon can be
understood in terms of the problem of diffusion in quenched random
potentials~\cite{szendro08}.

Recently, the evolution of {\em infinitesimal} perturbations in spatially extended chaotic
systems has been shown to be generically described by Langevin-type equations with
multiplicative noise~\cite{pik94,pik98,szendro08,lopez,sanchez}. A remarkable observation
in many systems~\cite{pik98} is that, after a suitable logarithmic transformation, the
statistical description of the dynamics of perturbations is captured by the prototypical
stochastic surface growth equation of Kardar-Parisi-Zhang (KPZ)~\cite{kpz}. In the surface
picture erratic fluctuations, due to the chaotic nature of the trajectory, are treated as
an effective noise. The existence of short-range correlations, coming from the
deterministic nature of the trajectory, are irrelevant for the scaling description of the
surface statistics. It is only natural that the existence of long-range
correlations~\cite{pik01} or a fat tail noise~\cite{sanchez} may change the universality
class observed. The surface picture has also been shown to be very useful to deal with the
dynamics of {\em finite} perturbations in the presence of
STC~\cite{lopez,primo05,primo06}.

In view of the successful description of the main LV as a nonequilibrium rough surface a
question that naturally arises is to what extent can we describe LVs corresponding to
other unstable directions in terms of surface roughening processes?
This is precisely the question we recently addressed in a Rapid communication \cite{szendro07},
and that we develop here for a variety of systems.

In the existing literature one finds that the LVs are commonly defined as the vectors that
appear as a byproduct of the standard Gram-Schmidt orthonormalization procedure to obtain
the LEs. This is largely due to the popularity of Benettin's
algorithm~\cite{benettin80,wolf85} to compute the Lyapunov spectrum in all kind of
dynamical systems. However, these vectors do {\em not} point in the most unstable
directions, but are forced to form an orthogonal set. This is not a minor point because
these vectors lack the intended physical meaning, which ultimately renders the
Gram-Schmidt vectors useless for many purposes. For example, when the $n$th Gram-Schmidt
LV, $\bm{e}_n(t)$, is left to evolve freely it will {\em not} grow exponentially with its
associated LE $\lambda_n$ (apart from the case $n=1$); instead, $\bm{e}_n(t)$ will
generally collapse in the direction of the first LV. However if the same vector
$\bm{e}_n(t)$ is integrated backwards in time it will shrink with exponent $-\lambda_n$
(neglecting numerical round-off errors). Not less important is the fact that Gram-Schmidt
LVs depend on the scalar product convention used because it defines the
orthogonalization condition.

These important caveats have attracted renewed interest in the problem of finding the
correct set of vectors that carry the dynamical information in systems exhibiting STC.
Recent work has focused on the properties of a different set of
vectors~\cite{wolfe_tellus07,szendro07,ginelli07}, the characteristic LVs (also called
covariant LVs), that are invariant under time reversal and covariant with the dynamics.
This vector set is independent of the scalar product used and provides an intrinsic
decomposition in tangent space which should correspond exactly with Oseledec's
splitting~\cite{eckmann}. Although the existence of the characteristic LVs is known since
long~\cite{eckmann,legras96,politi98,trevisan98} it has not been until recent times that efficient
algorithms have been devised to compute them~\cite{wolfe_tellus07,ginelli07}.

We have recently shown in Ref.~\cite{szendro07} that characteristic LVs carry important
information about the real-space structure, localization properties and space-time
correlations, which can be put in the form of a dynamical scaling of the associated rough
surfaces. These scaling properties were demonstrated for the particular case of 
lattices of coupled logistic maps,
but conjectured to be valid for a wide range of systems (at least
including all those reported in Ref.~\cite{pik98} as belonging to the KPZ class).

In this paper we study the spatiotemporal structure of the characteristic LVs in different
model systems exhibiting STC. Our aim is to analyze the spatial structure of the
characteristic LVs. In particular, we wish to provide further verification of the
previously reported scale-invariant properties of the LV surfaces (to be defined below)
and its validity for systems that differ significatively from the special case of
coupled-map lattices. Here we analyze systems of very different nature, including a
coupled-map lattice (CML), the (continuous-time) Lorenz `96 model, and a minimal
stochastic partial differential equation (PDE). We show that the leading LVs
(corresponding to the largest LEs) generically exhibit scale-invariant properties
inherited from those of the main vector. Our present results confirm and extend our
earlier claim~\cite{szendro07} concerning the generic, model independent, scaling
properties of characteristic LVs corresponding to unstable intrinsic directions.

\section{Models of Spatiotemporal chaos}
\label{sec_models}
Three spatially extended systems are studied in this paper: a coupled-map lattice, a
continuous-time model, and a stochastic equation. These models cover a range of dynamical
systems of very different nature, including discrete and continuous systems. We study
models that exhibit STC for a range of parameters. Since the scaling properties that we
are interested in are independent of microscopic details, our intention here is not to be
exhaustive in the exploration of model parameters or different terms in a particular
model, which have no effect on the scaling properties whatsoever. On the contrary, our aim
is to address much more generic types of models, such as those that are discrete 
or continuous in space or time.

\subsection{Coupled-map lattices}
\label{sec_cml}
Coupled-map lattices are simple prototypes of STC at low computational cost~\cite{bohr}.
This ultimately explains their widespread use to study different aspects of STC, which
would be prohibitively demanding in computation time should PDEs be used, for instance.
Here, we consider a ring of $L$ maps with diffusive coupling
\begin{eqnarray}
&u_i(t+1)= \epsilon f(u_{i+1}(t))+\epsilon
f(u_{i-1}(t)) + & \nonumber\\ 
& + (1 - 2\epsilon )f(u_i(t)),&
\label{cml}
\end{eqnarray}
where $\epsilon$ is the coupling parameter and $f$ is a map with chaotic dynamics.
Infinitesimal random perturbations evolve in tangent space following the linear equation
\begin{eqnarray}
&\delta u_i(t+1) = \epsilon f^{\prime}(u_{i+1}(t))\delta u_{i+1}(t) +& \nonumber\\ 
& + \epsilon f^{\prime}(u_{i-1}(t))\delta u_{i-1}(t) +
(1-2\epsilon)f^{\prime}(u_i(t))\delta u_i(t),&
\label{cml_lv}
\end{eqnarray}
where $f^\prime(\varrho)$ is just the derivative of the map $f(\varrho)$ with respect to
its argument $\varrho$. We have recently reported in Ref.~\cite{szendro07} about our study
of the space-time structure of LVs in the case of the lattice of coupled logistic maps
$f(\varrho)=4\varrho (1-\varrho)$. Here, as a further example we include the study of a
different type of map. The results we report on in this paper (see below) are analogous to
those obtained  for  logistic maps and thus we may conclude that no important differences
should arise among one-dimensional CMLs composed of continuous chaotic maps in one variable. 
Throughout this paper, we consider the skew tent map
with the same parameters as those in Ref.~\cite{ginelli07}:
\begin{eqnarray}
f(\varrho)=\left\{
\begin{array}{cc}
a \, \varrho     \qquad \qquad \qquad (\varrho \le 1/a)\\
a \, (\varrho-1)/(1-a) \qquad  (\varrho > 1/a)
\end{array}
\right. 
\end{eqnarray}
with $a=2.3$. The Lyapunov spectrum of the CML in Eq.~(\ref{cml}) for a coupling strength
$\epsilon=0.2$ is shown in Fig.~\ref{fig1}(a).

\subsection{Lorenz `96 model}
\label{sec_l96}
The second model we consider in this paper is an example of a chaotic continuous-time
system. This model is in many aspects very different from a CML model due to the
continuous character of the time variable. We study the model proposed by Lorenz in
1996~\cite{lorenz96} as a toy model in the context of weather dynamics. We consider the
variables $y_i$ defined in a ring, $i = 1, \cdots, L$, and the evolution equations
\begin{equation} 
\frac{d}{dt}y_i = -y_i-y_{i-1} (y_{i-2}-y_{i+1}) + F.
\label{l96} 
\end{equation}
The variables $y_i$ may be looked at as the values of some unspecified scalar
meteorological observable, like a vorticity or temperature, at equally spaced sites
extending around a latitude circle~\cite{lorenz_emanuel}. The model contains
quadratic, linear and constant terms mimicking advection, dissipation and external
forcing, respectively.

An infinitesimal perturbation $\delta y_i(t)$ evolves in tangent space according to the
linearized dynamics
\begin{eqnarray} 
\frac{d}{dt}\delta y_i & = -\delta y_i - (y_{i-2}-y_{i+1}) \delta y_{i-1} - y_{i-1} \delta
y_{i-2}
+ & \nonumber\\
& + y_{i-1} \delta y_{i+1},&
\label{l96_lv} 
\end{eqnarray}
which also governs the dynamics of any characteristic LV, as they are freely evolving
covariant perturbations.

\begin{figure}
 \centerline{\includegraphics *[width=75mm]{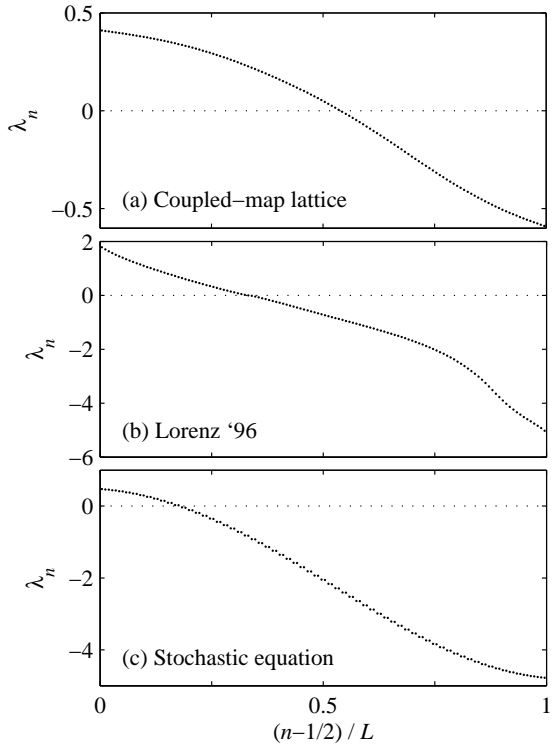}}
\caption{Lyapunov spectra for the three models studied in this paper. We follow the
standard convention and sort LEs in decreasing order $\lambda_1 \geq \lambda_2 \geq \cdots
\geq \lambda_L$. (The LEs were obtained for $L=128$, larger systems yield Lyapunov spectra
that overlap these ones.)
\label{fig1}}
\end{figure}

Regardless of how well or poor Eq.~(\ref{l96}) represents the
atmosphere, the model is nowadays an essential tool in studies of weather dynamics as a
testbed for forecasting techniques like breeding or singular
vectors~\cite{lorenz96,lorenz_emanuel,boffeta}. For $F=8$ the model exhibits STC, as
demonstrated by computing the Lyapunov spectrum shown in Fig.~\ref{fig1}(b). A Runge-Kutta
scheme is usually recommended for the numerical integration of Eq.~(\ref{l96}) to avoid
numerical instabilities. We have used a fourth order Runge-Kutta integration algorithm
with time step $\Delta t= 10^{-2}$, while to achieve the same precision with the Euler
method a much smaller time step, $\Delta t=1.5 \times 10^{-4}$, was needed.

\subsection{Multiplicative stochastic equation}
\label{sec_mse}
The third model we study is a multiplicative stochastic equation, which mimics the linear
evolution of infinitesimal perturbations in tangent space for spatio-temporal chaotic
systems. Pikovsky and Politi proposed~\cite{pik98} this stochastic PDE as the proper
candidate for modeling the statistical features of the dynamics of freely evolving
perturbations. Therefore, the analysis of this model will show to what extent the observed
scaling of characteristic LV surfaces is generic and model-independent in the context of
STC.

We consider a perturbation $\phi(x,t)$, initially homogeneous and random, whose time
evolution can be described, in a statistical sense, by the multiplicative Langevin
equation
\begin{equation}
\partial_t \phi = \zeta(x,t)\, \phi + \partial_{xx} \phi,
\label{mse}
\end{equation}
where $\zeta$ is a noise term that accounts for the chaotic fluctuations and one can
simply assume it to be Gaussian and uncorrelated: $\left< \zeta(x,t) \, \zeta(x',t')
\right>=2 \sigma \, \delta(x-x') \, \delta(t-t')$. It is worth stressing here that the
presence of short-range correlations in the noise term $\zeta$ (due to the purely
deterministic nature of the fluctuations), is actually irrelevant for the statistical
description in the long-wavelength limit, as already shown in the original work of
Pikovsky and Politi~\cite{pik98}.

We have numerically integrated Eq.~(\ref{mse}) by a stochastic Euler scheme (the noise
term up to order $\Delta t$) with a space and time step $\Delta x=1$ and $\Delta t=
10^{-2}$. The Lyapunov spectrum was computed and averaged over different noise
realizations (equivalent to different trajectories). In Fig.~\ref{fig1}(c) we plot the LEs
for a noise amplitude $\sigma=0.5$.

The multiplicative Langevin equation~(\ref{mse}) can be seen as a stochastic field theory
for the evolution of random errors in extended homogeneous systems. This stochastic model
has been found to describe the statistical properties of perturbations in many dynamical
systems ranging from lattices of logistic, tent or symplectic maps to the complex
Ginzburg-Landau equation~\cite{pik94,pik98,pik01}. It has also been extended to construct
a stochastic field theory of chaotic synchronization of extended
systems~\cite{ahlers02,mamunoz03}. Very recently, it has also been shown that a version of
Eq.~(\ref{mse}), which includes quenched disorder terms, describes the propagation of
perturbations in inhomogeneous chaotic systems~\cite{szendro08}.

Interestingly, the application of the Hopf-Cole transformation, $h=\ln|\phi|$,
immediately maps the problem into the KPZ equation for surface growth:
\begin{equation}
\partial_t h = \zeta + (\partial_x h)^2 + \partial_{xx} h,
\label{kpz}
\end{equation}
which ultimately justifies why the log-transformed (main) Lyapunov vector of many
spatiotemporal chaotic systems is found to belong to the KPZ universality
class~\cite{pik94,pik98}. 

There is an interesting caveat concerning this mapping, which has not been noticed before
in the context of STC. One can see that Eq.~(\ref{mse}) is invariant under the sign change
of the field $\phi \to - \phi$. However, the solutions of Eq.~(\ref{mse}) actually exhibit
a spontaneous breaking of this essential symmetry. In our numerical integration we observe
that for any random initial condition, no matter the spatial distribution of signs
for the initial field $\phi(x,t=0)$, with probability one the solutions of
Eq.~(\ref{mse}) asymptotically become either strictly positive or negative, ({\it i.~e.},
for long enough times $\phi(x,t) \neq 0$ for all $x$). The reason for this symmetry
breaking can be traced back to the mathematical properties of Eq.~(\ref{mse}). The key
observation is that the dynamics governed by Eq.~(\ref{mse}) cannot produce new zeros of
the field $\phi$. Therefore, sites where $\phi$ changes sign can only diffuse in the $x$
axis and, in the event two $\phi$-zeros collide, disappear. As we will see later on the
annihilation of zeros is crucial to understand the spatial structure of characteristic
LVs.

\section{Lyapunov vectors}
\label{sec_lvs}
In short, LVs are defined as the vectors
in tangent space that point towards the
directions such that a given perturbation expands (shrinks) with the corresponding LE.
Their physical significance arises from Oseledec's theorem~\cite{oseledec}. Let us now
discuss the definition and physical meaning of backward, forward, and characteristic LVs. 

Consider a nonlinear dynamical system governed by
\begin{equation}
\frac{d}{dt} \bm{u}(t) = {\mathbf \Upsilon}[\bm{u}]
\end{equation}
where $\bm{u}(t) \in {\mathbb R}^L$ is the position of the system in phase space and
${\mathbf \Upsilon}: {\mathbb R}^L \to {\mathbb R}^L$ is the nonlinear evolution operator.
Infinitesimal perturbations $\bm{\delta u}(t)$ follow the linear dynamics given by the
tangent space equations:
\begin{equation}
\frac{d}{dt}\bm{\delta u}(t) = \frac{\partial {\mathbf \Upsilon}[\bm{u}]}
{\partial \bm{u}} \, \bm{\delta u}(t),
\label{pert}
\end{equation}
which implies that the perturbation can be computed at time $t$ from the perturbation at
an earlier time $t_0$ as
\begin{equation}
\bm{\delta u}(t) = {\mathbf M}(t,t_0) \bm{\delta u}(t_0),
\end{equation}
where ${\mathbf M}(t_0,t) = {\mathbf M}(t,t_0)^{-1}$ is some linear operator.

\subsection{Backward (and forward) Lyapunov vectors}
\label{sec_blvs}
According to Oseledec's theorem~\cite{oseledec} (details can also be found in
Ref.~\cite{eckmann}) there exists the remote past limit symmetric operator ${\mathbf
\Phi}_{\mathrm b}(t) = \lim_{t_0 \to -\infty} [{\mathbf M}(t,t_0) {\mathbf
M}^*(t,t_0)]^{1/[2(t-t_0)]}$, where ${\mathbf M}^*$ is the adjoint operator. All $L$
eigenvalues of ${\mathbf \Phi}_{\mathrm b}(t)$ are positive time-independent numbers that
can be written as $\exp (\lambda_n)$, where $\lambda_n$ are the LEs, and the corresponding
eigenvectors form an orthonormal basis $\{\bm{e}_n(t)\}$, $n=1, \cdots, L$. These
eigenvectors are called {\em backward} LVs~\cite{legras96} and represent the directions in
tangent space that, at the present time $t$, are seen to have grown at exponential rates
$\lambda_n$ since the remote past. The set of backward LVs is precisely the orthonormal
set obtained using the standard Gram-Schmidt orthogonalization method to compute the
LEs~\cite{ershov98}. 

Conversely, {\em forward} LVs form a different orthonormal set of vectors analogous to
backward LVs, but with the temporal properties inverted. In this case they are obtained as
the eigenvectors of the far future limit operator $\lim_{t_0 \to \infty} [{\mathbf
M}^*(t_0,t){\mathbf M}(t_0,t)]^{1/[2(t_0-t)]}$, which obviously has the same eigenvalues
as ${\mathbf \Phi}_{\mathrm b}(t)$. When left to evolve freely from the present time $t$, the
$n$th forward LV grows exponentially in the far future at a rate given by the
corresponding LE $\lambda_n$. However under reverse (time backwards) integration all
forward LVs collapse into the last forward LV.

The popularity of the algorithm of Benettin {\em et al.}~\cite{benettin80,wolf85} for
computing the first $n$ Lyapunov exponents, via successive Gram-Schmidt orthonormalization
of a set of $n$ vectors that evolve according to the linear equations in tangent space,
has caused many authors to consider using the resulting orthonormal set $\{\bm{e}_n(t)\}$,
$n=1, \cdots, L$, as the Lyapunov vectors. As mentioned in the Introduction, the use
of this set of vectors poses serious problems in certain applications. Any of the $L$
backward LV tends to align exponentially fast with the first LV. This has to be avoided
by the externally imposed orthogonalization, which `resets' the vector set every few time
steps.
Moreover, different scalar products produce different sets of backward and forward LVs.

\subsection{Characteristic Lyapunov vectors}
\label{clvs}
In order to construct a complete set of $L$ characteristic (or covariant) vectors,
$\{\bm{g}_n(t)\}$, $n=1, \cdots, L$, independent of the scalar product and having the
wanted topological properties, one has to intersect the subspaces spanned by the backward
and forward LV in a precise manner as discussed by Eckmann and Ruelle~\cite{eckmann}. 

At variance with backward and forward LVs, characteristic vectors have the desired
topological and dynamical properties: (i) They are independent of the scalar product; (ii)
They reduce to the Floquet eigenvectors for a periodic orbit~\cite{trevisan98}; (iii) Any
given $\bm{g}_n(t)$ grows at an exponential rate given by the associated LE $\lambda_n$ in
the far future, and with rate $-\lambda_n$ backward-integrating to the remote past (under
the linearized equations in tangent space, with no orthogonalization or any other external
constraint). For instance, in chaotic continuous-time systems, and in contrast with
backward LVs, there is a characteristic LV tangent to the trajectory that corresponds to
the zero LE associated with time-shift invariance.

Contrary to the (artificial) orthogonal disposition of backward LVs, characteristic LVs
generally do not form an orthogonal set. We note that the first backward and
characteristic LVs are tangent, $\bm{g}_1(t) \propto \bm{e}_1(t)$. For $n>1$, the $n$th
characteristic LV is a linear combination of backward LVs from $1$ to $n$.

Although Eckmann and Ruelle~\cite{eckmann} already discussed these ideas in 1985, they had
received little attention in the literature until very recently. This is partially due to
the fact that implementing such a theoretical construction is by no means a simple task
from a computational point of view. Only very recently, Wolfe and
Samelson~\cite{wolfe_tellus07} have proposed a computationally efficient algorithm to
obtain the set of characteristic LVs.
We have used this algorithm in all our calculations, and
technical details can be found in the Appendix.
Also Ginelli {\it et. al.} have proposed a similar algorithm~\cite{ginelli07}.

In the rest of this paper we study the spatial structure of LVs, focusing on universal
features that are shared among different models of STC.

\section{Surface growth picture}
\label{sec_surf}
In systems with spatiotemporal chaos the first LV localizes in space, so that its
magnitude spans several orders of magnitude between the top and the valleys. In
homogeneous systems, translational invariance implies that the localization site is not
static.

It was recognized some time ago~\cite{pik94,pik98} that the spatiotemporal dynamics of the
first LV is much more understandable as a surface to be obtained after Hopf-Cole
transforming the vector. Until recently, very little
was known about the spatial correlations of
characteristic (or backward) vectors for $n > 1$. We have
reported~\cite{szendro07} on the existence of intrinsic length scales and have determined
the form of the spatiotemporal correlations of LVs corresponding to the leading unstable
directions by translating the problem to the language of scale-invariant growing surfaces.
We found that characteristic LVs corresponding to the most unstable directions also
exhibit spatial localization, strong clustering around given spatiotemporal loci, and
remarkable dynamic scaling properties of the corresponding surfaces. In contrast, any two
backward LVs localize in different sites since they are mutually orthogonal. Also, they do
not exhibit dynamic scaling due to artifacts in the dynamical correlations by
construction~\cite{szendro07}. Our results were based on numerical studies of lattices of
coupled-maps, but conjectured to be generically valid for a wide range of systems. Our aim
here is to extend our previous analysis and put it in a wider context. For this purpose,
the analysis of the stochastic model Eq.~(\ref{mse}) has a particular significance.

\begin{figure}
\centerline{\includegraphics *[width=75mm]{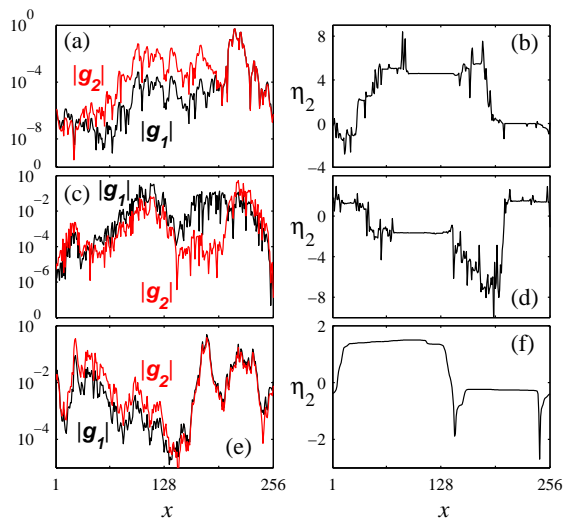}}
\caption{(Color online) The first and the second characteristic Lyapunov vectors for: (a)
Coupled-map lattice (\ref{cml}), (c) Lorenz `96 model, and (e) Stochastic equation
(\ref{mse}). Note that we take the absolute values and a logarithmic scale for the y-axis.
Panels (b),(d), and (f) show the fields $\eta_2=h_2-h_1$ for the three models.\label{fig2}}
\end{figure}

Figures~\ref{fig2}(a), \ref{fig2}(c), and \ref{fig2}(e) show typical snapshots of
the first and second characteristic LVs in logarithmic scale for the three models
introduced in Sec.~\ref{sec_models}. One can see that both vectors may localize in
the same site (which is not possible for backward LVs due to their mutual orthogonality).
For every characteristic LV, $\bm{g}_n(t) = [g_n(x,t)]_{x=1}^{x=L}$, we define a surface
via the Hopf-Cole transformation, $h_n(x,t)=\ln |g_n(x,t)|$. For the sake of comparison we
will also consider the surfaces associated with backward LVs: $h_n(x,t)=\ln |e_n(x,t)|$.
After the mapping the $n$th LE corresponds to the average velocity of the corresponding
$n$th surface, $\langle (1/L) \sum_{x=1}^{x=L} h_n(x,t) \rangle = \langle \ln
\prod_{x=1}^{x=L} |g_n(x,t)|^{1/L} \rangle \approx \lambda_n t$. 

The surface growth formalism has allowed to identify different universality classes in
spatiotemporal chaotic systems~\cite{pik94,pik98,pik01,sanchez}. In particular, the
universality class of KPZ has been widely observed in non-Hamiltonian systems with no
special conservation laws, discontinuities or broken symmetries. This includes, among
others, lattices of coupled logistic maps, and the Ginzburg-Landau and
Kuramoto-Sivashinsky PDEs. The three model systems studied in this paper (see
Sec.~\ref{sec_models}) also belong to KPZ universality class. This can be confirmed by
calculating the so-called dynamic and roughness exponents.

Interestingly, we have found that the $n$th LV (either characteristic or backward) is a
piecewise copy of the main LV. This behavior is conveniently captured by the
difference-field $\eta_n \equiv h_n-h_1$. For instance in Figs.~\ref{fig2}(b),
\ref{fig2}(d), and \ref{fig2}(f) we plot the difference-field $\eta_2=h_2-h_1$, which
reveals the existence of flat regions indicating that the first and the second LV surfaces
are strongly correlated. The second LV surface is loosely speaking ``piecewise KPZ", since
it is made of pieces that differ from the main vector at only a few sites. Actually, the
$n$th LV also exhibits the same structure for increasing $n$-- namely, the
difference-field $\eta_n$ is also formed by flat regions separated by fluctuating edges.
The typical plateau length of the field $\eta_n$ naturally defines a characteristic
length scale $\ell_n$, below which the $n$th surface is identical to the first surface.
This characteristic plateau size decreases with increasing $n$. 
So that, beyond some
$n_{\rm max}$, the number of fluctuating edges is so large ($\ell_n \to 1$ for $n \gg
n_{\rm max}$) that the ``piecewise KPZ" picture is not useful any longer.

It is remarkable that, for systems whose first LV belongs to the KPZ class, there is a
finite part of the Lyapunov spectrum 
($\lambda_n$ with $n<n_{\rm max}$) that can be understood in
terms of piecewise copies of the first vector. Note that this peculiar spatial structure
can only be easily identified after the logarithmic transform.
Last but not least, this spatial structure is also consistent with the fact that,
any characteristic LV is governed by the same tangent dynamics for a given system,
Eqs.~(\ref{cml_lv}), (\ref{l96_lv}), and (\ref{mse}), and a given trajectory.

\section{Spatial structure}
\label{sec_struct}
\begin{figure}
\centerline{\includegraphics *[width=75mm]{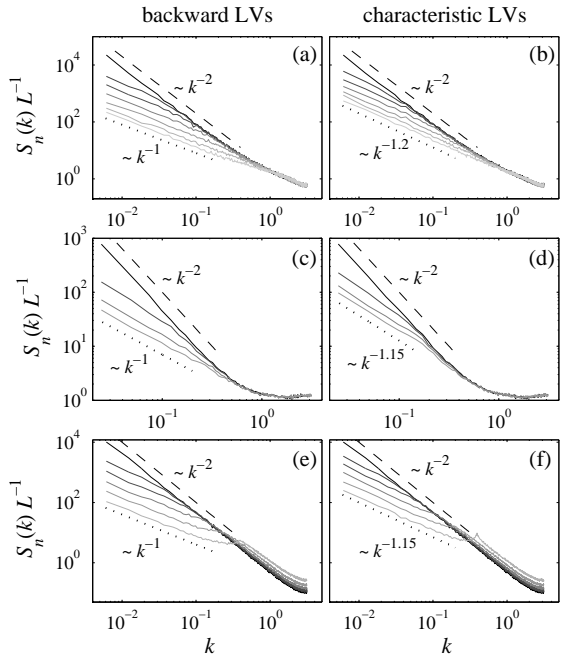}}
\caption{Structure factors for LVs of the three models considered in this paper: (a,b)
Coupled-map lattice (\ref{cml}) with $L=1024$, from top to bottom $n=1,4,8,16,32,64,128$;
(c,d) Lorenz `96 model (\ref{l96}) with $L=256$, $n=1,4,8,12$; and (e,f) Stochastic
equation (\ref{mse}) with $L=1024$, $n=1,4,8,16,32,64$. In all cases backward LVs (a,c,e)
beyond the first one decay at small wavelengths as $k^{-1}$. Characteristic LVs (b,d,f)
display stronger correlations with $k^{-\gamma}$ ($\gamma \approx 1.15$-$1.2$).
We averaged over 200 realizations for the CML and 1000 for the other two models.}
\label{fig3}
\end{figure}

In this section we carry out a quantitative description of the spatial correlations of the
LV surfaces $h_n(x,t)$. We compute the stationary structure factor $S_n(k)= \lim_{t \to
\infty} \langle \hat h_n(k,t) \hat h_n(-k,t) \rangle$, where $\hat h_n(k,t)= \sum_x
\exp(ikx) h_n(x,t)$, and the brackets indicate an average over different system
trajectories (or noise realizations in the case of the purely stochastic model). As
expected the first LV surface correlations decay as $k^{-2}$ (Fig.~\ref{fig3}), in
agreement with KPZ universality class~\cite{kpz,pik98}. Interestingly, the $n$th LV
surface for $n > 1$ also shows scale-invariant correlations $\sim k^{-2}$, with a
crossover to a different scaling regime at a wavenumber $\Bbbk_n$ that depends on $n$. It
is natural to link this crossover length scale to the plateaus discussed in the preceding
section. Indeed, we have shown in Ref.~\cite{szendro07} that this crossover wavelength is
related to the typical size of plateaus $\Bbbk_n\sim \ell_n^{-1}$.

At long wavelengths, correlations of LV surfaces associated with backward and
characteristic LVs decay approximately as $k^{-1}$ and $k^{-1.2}$, respectively
(Fig.~\ref{fig3}). This $1/k$-divergence indicates extremely weak long-range spatial
correlations for both classes of LVs. However, backward and characteristic vectors exhibit
markedly different dynamical properties. 
To be precise, the imposition of orthogonality causes the mapping of a
backward LV at $t$ into itself at $t+\Delta t$ to convey $1/k$ long-range correlations.
On the contrary,
characteristic LVs show increasing correlation lengths as time evolves, as one would expect 
for a surface evolving with local equations.
In this case (and in contrast with backward LVs), 
surface correlations are found to satisfy dynamic scaling
akin to growing surfaces (cf.~Fig.~4 in \cite{szendro07}).

Deterministic equations (\ref{cml}) and (\ref{l96}) yield LVs whose spatial structure is
analogous to the structure of LVs obtained with the stochastic equation (\ref{mse}) with
white noise. This indicates that in spatiotemporal chaotic systems of the KPZ universality
class, the role of spatio-temporal correlations is insignificant in what concerns the
statistical (long-time and large-scale) structure of LVs.

Finally, we recall that when $n$ becomes large ($n>n_{\rm max}$) specific features of each
model will show up. For instance, in the case of the multiplicative stochastic equation,
the $n$th LV appears as a noisy sinusoidal function because diffusion prevails over the
stochastic term. Accordingly a peak appears at intermediate wave numbers in the structure
factor [see the curve for $n=64$ in Fig.~\ref{fig3}(f)].

\section{Multiplicative stochastic equation}
\label{sec_mult}
The multiplicative Langevin model discussed in Sec.~\ref{sec_mse} constitutes a minimal
model for describing the dynamics of free perturbations in a (wide) family of systems
exhibiting spatiotemporal chaos~\cite{pik98}. In particular, since random free
perturbations rapidly tend to be tangent to the main LV, Eq.~(\ref{mse}) also describes
the scaling behavior of the first LV. As we have discussed in the preceding section,
characteristic LVs are freely propagating perturbations, covariant with the dynamics as
well as with the time inverted dynamics. Therefore, we conjecture here that the
multiplicative stochastic model should also describe the statistics and scaling behavior
of the $n$th characteristic LV, at least for $n < n_{\rm max}$. In this regard, the
scaling properties of the $n$th characteristic LV in systems with STC should be
generically linked to those of the solutions supported by the multiplicative Langevin
model. In this section, we study in more detail the structure of the solutions and LVs in the
multiplicative stochastic model.

\begin{figure}
\centerline{\includegraphics *[width=75mm]{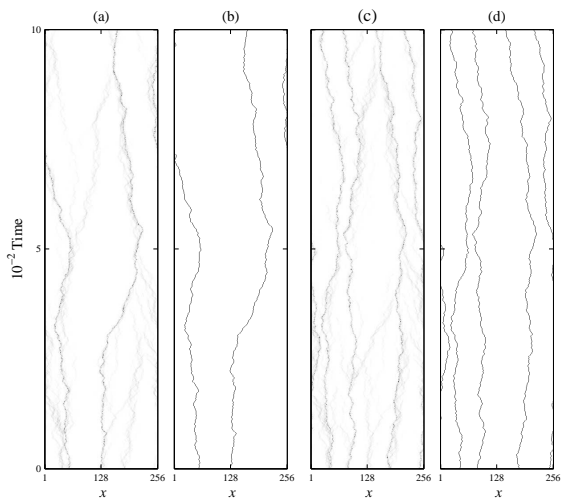}}
\caption{Spatiotemporal plot of the fields (a) $|\partial_x \eta_2|$ and (c) $|\partial_x
\eta_4|$ for the stochastic PDE ($\ref{mse}$) with $L=256$; the plateaus appear as
the clear regions. 
The darkest regions correspond to zeros of vectors ${\bm{g}_2}$
and ${\bm{g}_4}$, indicated in panels (b) and (d), respectively.
Other light gray regions in (a) and (c) correspond to kinks discussed in the text.}
\label{time}
\end{figure}

We have computed the characteristic LVs for the stochastic model. Figure~\ref{time}
demonstrates the existence of plateaus for the differences $\eta_n(x,t)=h_n-h_1$, whose size
decreases with $n$. The plateaus are bounded by kinks, which are prominently placed at the
sites where $g_n(x,t)$ crosses zero ({\it i.e.} where $h_n(x,t) \rightarrow -\infty$).

The asymptotic attracting solution of Eq.~(\ref{mse}) is the first LV, $\phi(x,t) =
g_1(x,t)$. As discussed in Sec.~\ref{sec_mse}, the asymptotic solution $g_1(x,t)$ has the
same sign everywhere. This solution is univocally determined for a given trajectory (noise
realization), apart from an arbitrary non-zero constant factor. The solution $\phi(x,t) =
g_1(x,t)$ has the statistical properties of a KPZ surface because the Hopf-Cole
transformation from Eq.~(\ref{mse}) to Eq.~(\ref{kpz}) is exact for $n=1$. In contrast,
characteristic LVs for $n>1$ are saddle solutions of Eq.~(\ref{mse}), which are forced to
have regions with opposite signs. This, in turn, naturally leads to smaller growth rates
($\lambda_n<\lambda_1$).

We find that the number of zeros of the $n$th LV is $\mathcal{N}_0(n)= 2
\left[n/2\right]$, where $[q]$ stands for the integer part of $q$. Note that
Eq.~(\ref{mse}) is not able to create new zero crossings, which implies that
$\mathcal{N}_0(n)$ cannot fluctuate and is a conserved quantity. We also remark that
$\mathcal{N}_0(n)$ corresponds to the number of zeros of the $n$th normal mode of the
(noise-free) diffusion equation $\partial_t \phi = \partial_{xx} \phi$, assuming they are
ordered according to their stability.

It would be very interesting to be able to write down the stochastic PDE describing the
dynamics of the surface associated with the $n$th characteristic LV. However, this turns
out to be a very difficult task. A more qualitative description can nonetheless be very useful.
The first LV ($n=1$) has no zeros and, as mentioned
above, this allows us to exactly transform Eq.~(\ref{mse}) into the KPZ equation
(\ref{kpz}). However for $n>1$, each $g_n(x,t)$ has $\mathcal{N}_0(n)$ zeros, which
cannot be neglected when applying the Hopf-Cole transform. Indeed, one can observe that
$\lambda_n = \langle \overline{\partial_t h_n} \rangle \neq \langle \overline{(\partial_x
h_n)^2} \rangle$, which indicates that there must be other terms contributing to the
velocity of the $n$th Lyapunov surface. As expected, the equality $\lambda_1 =
\langle\overline{\partial_t h_1} \rangle = \langle \overline{(\partial_x h_1)^2} \rangle$
exactly holds in the singularity-free case $n=1$.

A detailed analysis using the discrete version of (\ref{mse}) reveals that the stochastic
PDE governing $h_n$ is a KPZ equation with singular (and difficult-to-treat) terms at the
points where $h_n \rightarrow -\infty$ ($g_n \rightarrow 0$). Formally one can expect to
have
\begin{equation}
\partial_t h_n =  \zeta + (\partial_x h_n)^2 + \partial_{xx} h_n 
+ \sum_{i=1}^{\mathcal{N}_0(n)} \Xi[x_i(t)],
\label{kpzst}
\end{equation}
where the function $\Xi[x_i(t)]$ accounts for singular delta-like contributions at the
zeros $x_i(t)$, whose positions move erratically around the system.

We first note that the
erratic motions of the zeros [Figs.~\ref{time}(b) and \ref{time}(d)] seem not
to be the source of the long-ranged correlations. At long times, the erratic motion of
zeros is sub-diffusive: the position of the $i$th zero satisfies
$\langle\left(x_i(t)-x_i(0)\right)^2\rangle \sim t^\gamma$, and we find
$\gamma\approx 0.87$ for the second LV and $\gamma\approx 0.62$ for the fourth LV from
numerical simulations in a system of size $L=256$. We have already shown~\cite{szendro07}
that at long wavelengths (in the $S(k) \sim k^{-1.2}$ region) surface correlations of
characteristic LVs exhibit dynamic scaling. The
analysis of coupled-map lattices shows a fast propagation of correlations at large scales
with a dynamic exponent $z=1$ corresponding to a ballistic process ($\gamma= 2/z = 2$).
Since zeros do not propagate ballistically, but sub-diffusively, we conclude that
information propagation at long wavelengths is mediated by a different process. The best
candidates are small kinks [see, for instance, a typical kink at $x \approx 100$ in
Fig.~\ref{fig2}(f)] that can be identified (light gray traces in the plots) as traveling
objects in Figs.~\ref{time}(a) and \ref{time}(c). Interestingly, the dynamics of the kinks
is governed by the equation of the field $\eta_n$, which can be written exactly inside
a plateau region:
\begin{equation}
\label{diego_eq}
\partial_t \eta_n = (\partial_x \eta_n)^2 + \partial_{xx}\eta_n +
2(\partial_x h_1)(\partial_x\eta_n).
\end{equation}
The drift term, proportional to $\partial_x\eta_n$, would lead to the ballistic dynamics
of the kinks with $z=1$. This provides the mechanism for the ballistic propagation of
correlations observed at long wavelengths.

\section{Discussion}

Our numerical results with the stochastic model (\ref{mse})
are particularly revealing since they explicitly show to what extent 
equations for growing surfaces can be used to describe STC.
Specifically, Eq.~(\ref{mse}) is invariant under multiplication by a constant, $\phi
\rightarrow c \phi$, which leads to the symmetry $h \rightarrow h + \ln c$ for the
corresponding surface. This symmetry property leads to scale invariance of
$h$~\cite{hents94}. As equations for infinitesimal perturbations are
always of linear type, Eq.~(\ref{pert}), the symmetry $\phi \rightarrow c \phi$
is always fulfilled and, in turn, systems with STC will exhibit scale invariance
of the associated Lyapunov vector surfaces. Different universality
classes, depending on the existence of 
correlations or conserved quantities, may be obtained.

A final remark is in order. The conservation of the number of zero crossings, observed for
Eq.~(\ref{mse}), is not fulfilled in general. In a generic setting, the dynamics of perturbations
would be governed by linear equations that might contain higher-order derivatives
multiplied by possibly fluctuating coefficients $\xi_i$: $\partial_t \delta u = \xi_1
\delta u + \xi_2 \partial_x\delta u + \xi_3 \partial_{xx} \delta u + \cdots$. Contrary to
the perhaps oversimplified stochastic model~(\ref{mse}), zeros can be created in this
general situation; for instance, if a drift term exists ($\xi_2\ne 0$), or if $\xi_3$ can
take negative values. Nevertheless, the scaling behavior of such a system is expected to
be correctly described by Eq.~(\ref{mse}), because those model-specific terms are actually
irrelevant in the sense of the renormalization group. In conclusion, the role of zeros is
very important to understand the dynamics of (\ref{mse}), but how they are linked to
structural properties of generic
systems with spatio-temporal chaos remains an issue for future work.

In summary, in this paper we have studied spatiotemporal chaos in three qualitatively
different (non-Hamiltonian) systems. In all cases 
characteristic (and backward) Lyapunov vectors exhibit very similar
spatial structure. 
The $n$th Hopf-Cole transformed LV is a piecewise copy of the first LV,
with a typical plateau length that decreases with $n$.
One of the three systems studied is a stochastic equation that 
serves as a minimal model for the leading LVs in systems
whose first LV belongs to the universality class of KPZ.

\acknowledgments
We thank A.~Pikovsky and A.~Politi for stimulating discussions.
Financial support from the Ministerio de Educaci\'on y Ciencia (Spain) under projects
FIS2006-12253-C06-04 and CGL2007-64387/CLI is acknowledged. D.P.~acknowledges support by
MEC (Spain) through the Juan de la Cierva Programme.

\section*{APPENDIX: COMPUTATION OF CHARACTERISTIC LYAPUNOV VECTORS}

In this Appendix we outline the procedure we have followed to obtain the characteristic
Lyapunov vectors. It is based on the work by Wolfe and Samelson \cite{wolfe_tellus07}. It
assumes that that there is no degeneracy in the Lyapunov spectrum; {\em i.e.}~$\lambda_1 >
\lambda_2 > \cdots > \lambda_L$. For the sake of concreteness we restrict the following
discussion to the CML model (\ref{cml}), but it is not difficult to extend it to
continuous-time systems.

Given the initial state of the system ${\bm u(t_0=0) = [u_1(t_0), u_2(t_0), \cdots,
u_L(t_0)]}$, infinitesimally small perturbations $\bm{\delta u}(t_0)$ in the
initial condition evolve up to linear order ({\it i.e.}~in tangent space) according to
\begin{eqnarray}
\delta u_i(t+1) &=& \epsilon f^{\prime}(u_{i+1}(t))\delta u_{i+1}(t)
+\epsilon f^{\prime}(u_{i-1}(t))\delta u_{i-1}(t) \nonumber\\  
&+& (1-2\epsilon)f^{\prime}(u_i(t))\delta u_i(t) \nonumber\\ 
&\equiv&\sum_{j=1}^L T_{ij}[{\bm u}(t)]\delta u_j(t), \nonumber
\label{tangent}
\end{eqnarray}
with $f^{\prime}$ being the derivative of $f(y)$ with respect to $y$ and
$\mathrm{\mathbf{T}}[{\bm u}(t)]$ the $L \times L$ Jacobian matrix evaluated at ${\bm
u}(t)$. The evolution of an infinitesimal perturbation $ \bm{\delta u}(t_1)$ is governed
by the linear equation: ${\bm {\delta
u}}(t_0)=\mathrm{\mathbf{M}}(t_0,t_1) {\bm{\delta u}}(t_1)$. The linear operator
$\mathrm{\mathbf{M}}$ is just the product of the Jacobian matrices evaluated along the
system trajectory from $t_1$ to $t_0$, {\it i.e.} 
\begin{equation}
\mathrm{\mathbf{M}} (t_0,t_1)\equiv
\mathrm{\mathbf{T}}[{\bm u}(t_0-1)]\mathrm{\mathbf{T}}[{\bm u}(t_0-2)]\ldots
\mathrm{\mathbf{T}}[{\bm u}(t_1+1)]\mathrm{\mathbf{T}}[{\bm u}(t_1)]
\nonumber
\end{equation}
According to Oseledec's theorem~\cite{oseledec} (details can be found in Ref.~\cite{eckmann})
there exists the limit operator 
\begin{equation}
{\mathbf \Phi}_{b}(t_0) = \lim_{t_1 \to -\infty}
[{\mathrm{\mathbf{M}}(t_0,t_1)} {\mathrm{\mathbf{M}}^*(t_0,t_1)}]^{\frac{1}{2(t_0-t_1)}}
\nonumber
\end{equation}
such that the logarithms of the eigenvalues are the LEs $\lambda_n$, and the eigenvectors
form an orthonormal basis $\{\bm{e}_n(t_0)\}$. This set of eigenvectors, so-called
{\em backward} LVs~\cite{legras96}, indicates the directions of growth of perturbations
grown since the remote past with exponents $\lambda_n$.
The backward LVs are precisely the orthonormal vectors
obtained using the standard Gram-Schmidt orthogonalization method to compute the
LEs~\cite{ershov98}.

Conversely the directions that will grow with exponents $\lambda_n$ are indicated by the
so-called forward Lyapunov vectors $\{\bm{f}_n(t_0)\}$. They constitute an orthogonal
basis 
formed by the eigenvectors of the 
operator: 
$${\bm \Phi}_{f}(t_0) = \lim_{t_2 \to \infty}
[{\mathrm{\mathbf{M}}^*(t_2,t_0)} {\mathrm{\mathbf{M}}(t_2,t_0)}]^{\frac{1}{2(t_2-t_0)}}
$$
As with the backward LVs, the Gram-Schmidt procedure
can be used to obtain forward LVs, but now going backwards in time 
and using the transposed Jacobian matrices because of the identity
$\mathrm{\mathbf{M}}(t_2,t_0)^*\equiv
\mathrm{\mathbf{T}^*}[{\bm u}(t_0)]\mathrm{\mathbf{T}^*}[{\bm u}(t_0+1)]\ldots
\mathrm{\mathbf{T}^*}[{\bm u}(t_2)]$. As noted by Legras and Vautard \cite{legras96},
the use of the transposed Jacobian (in contrast with the inverse ones) 
causes the forward LVs to come up with the standard ordering. This means that 
to obtain the first $n$ forward LVs we need to integrate only $n$ perturbations (instead
of $L-n+1$).
Finally, note that computing forward LVs requires to store a trajectory
${\bm u}(t_0),{\bm u}(t_0+1), \ldots, {\bm u}(t_2)$.

Each backward (resp.~forward) LV grows with its exponent $-\lambda_i$ (resp. $\lambda_i$)
when it is left 
to evolve in the limit
$t\rightarrow -\infty$ (resp. $t\rightarrow \infty$). However both sets, backward and
forward,
do not follow their associated exponents when the time limit is reversed.
For this reason it is much more interesting to consider another set of vectors 
$\{{\bm{g}_n(t)}\}$, the so-called {\em characteristic} LVs,
that grow with exponent $\lambda_n$ (-$\lambda_n$) when integrating to the far future
(past):
$$
\lim_{|t| \to \infty} (t-t_0)^{-1} \ln ||{\mathrm{\mathbf{M}}}(t,t_0) \bm{g}_n
(t_0) ||= \lambda_n. 
$$

The $n$th characteristic Lyapunov vector is a linear combination of the first $n$ 
backward Lyapunov vectors~\footnote{A component of ${\bm{g}_n}$ on ${\bm{e}_j}$ with $j>n$
would dominate the backward integration ($-\lambda_j > -\lambda_n$).}:
$$
{\bm{g}_n}=\sum_{i=1}^n \left<{\bm{e}_i},{\bm{g}_n}\right> {\bm{e}_i} 
\equiv \sum_{i=1}^n y_i^{(n)} {\bm{e}_i} 
$$
${\bm{g}_n}$ does not project on the subspace 
spanned by the $n-1$ first forward LVs. This allows,
by means of some algebraic transformation \cite{wolfe_tellus07}, 
to express
the coefficient vector ${\bm{y}^{(n)}}=(y_1^{(n)},y_2^{(n)},\ldots,y_n^{(n)})$ 
as the one-parametric family of nontrivial solutions of 
$$
{\mathrm{\mathbf{D}}}^{(n)} {\bm{y}^{(n)}} = {\bm{0}} ,
$$
where the $n\times n$
matrix ${\mathrm{\mathbf{D}}}^{(n)}$  is calculated using the first $n-1$ forward LVs:
$$
D_{kj}^{(n)}=\sum_{i=1}^{n-1} \left< \bm{e}_k , \bm{f}_i\right> \left< \bm{f}_i ,
\bm{e}_j\right> .
$$
${\bm{y}^{(n)}}$  is then completely determined (up to a global sign) imposing
normalization of ${\bm{g}_n}$.

\bibliographystyle{prsty}

\end{document}